# Subgroup Identification and Individualized Treatment Policies: A Tutorial on the Hybrid Two-Stage Workflow

Nan Miles Xi [1], Xin Huang [1], Lin Wang [2*]

[1] Data and Statistical Sciences, AbbVie Inc., North Chicago, IL 60064, USA

[2] Department of Statistics, Purdue University, West Lafayette, IN 47907, USA

* Correspondence: linwang@purdue.edu

## Abstract

Patients in clinical studies often exhibit heterogeneous treatment effect (HTE). Classical subgroup analyses provide inferential tools to test for effect modification, while modern machine learning methods estimate the Conditional Average Treatment Effect (CATE) to enable individual level prediction. Each paradigm has limitations: inference focused approaches may sacrifice predictive utility, and prediction focused approaches often lack statistical guarantees. We present a hybrid two-stage workflow that integrates these perspectives. Stage 1 applies statistical inference to test whether credible treatment effect heterogeneity exists with the protection against spurious findings. Stage 2 translates heterogeneity evidence into individualized treatment policies, evaluated by cross fitted doubly robust (DR) metrics with Neyman-Pearson (NP) constraints on harm. We illustrate the workflow with working examples based on simulated data and a real ACTG 175 HIV trial. This tutorial provides practical implementation checklists and discusses links to sponsor oriented HTE workflows, offering a transparent and auditable pathway from heterogeneity assessment to individualized treatment policies.

# 1 Introduction

In clinical studies, patients often exhibit substantial heterogeneity in treatment response. Some individuals derive strong benefit from the treatment, others experience little to no effect, and some may even be harmed [1]. Identifying these subgroups who benefit is critically important for decision making in precision medicine and health policy [2]. While randomized controlled trials (RCTs) traditionally focus on estimating the average treatment effect in the overall population, there is growing recognition that this average can mask important differences across subpopulations [3]. Identifying patients who benefit from an intervention can improve patient outcomes, avoid unnecessary treatments for those unlikely to benefit, and support personalized treatment strategies in practice [4].

There are two parallel tracks in heterogeneous treatment effect (HTE) analysis. One track focuses on population level inference, as in classical subgroup analysis of clinical trials [5]. Here the goal is to estimate treatment effects within clinically defined subgroups and test for differences. This paradigm provides p-values and confidence intervals to infer whether observed treatment effect differences are statistically significant. The other track emphasizes individual prediction and classification, as seen in modern machine learning approaches for Conditional Average Treatment Effect (CATE) estimation and uplift modeling [6,7]. These methods aim to predict whether an individual will benefit from treatment using high dimensional covariate information. Such techniques can uncover complex, multifactorial sources of heterogeneity that traditional subgroup analyses might miss. Yet, these two paradigms have not been fully integrated. Methods that prioritize interpretability and hypothesis testing often sacrifice predictive accuracy, whereas methods that excel at individual prediction often lack statistical inference guarantees. We propose a hybrid two-stage workflow to bridge this gap. This approach combines the strengths of both classical and machine learning methods, leveraging each in turn to ensure robust findings and practical utility.

The first stage of the workflow uses statistical testing or estimation with uncertainty quantification to assess whether HTE is present at the population level. Depending on the scientific question, this may involve an omnibus test for any treatment-covariate interaction, testing prespecified treatment-biomarker interactions, or formulating hypotheses about the existence of subgroups whose CATE exceeds a clinically meaningful margin $\delta$ [8,9]. By allocating the error budget across these options and clearly specifying the estimand and margin, Stage 1 provides auditable evidence at population level for or against meaningful effect modification and defines a gate for further analysis. When Stage 1 does not detect credible heterogeneity, the workflow stops and the negative finding is documented.

Conditional on passing this gate, the second stage translates heterogeneity evidence into an individualized treatment policy. We estimate CATE or benefit scores using flexible learners and evaluate treatment rules derived from these scores using cross fitted doubly robust (DR) pseudo-outcomes. Policy performance is summarized by uplift ranking metrics, the expected policy value relative to non-personalized strategies, and a Neyman-Pearson (NP) safety rule that constrains the proportion of treated patients whose true treatment effect falls below $\delta$. This separation between Stage 1 inference and Stage 2 decision making helps prevent exploratory subgroup searches from being conflated with per patient recommendation.

In the remainder of the paper, we develop this hybrid workflow as a tutorial. Section 2 introduces the conceptual framework, including target estimands, Stage 1 hypothesis options, Stage 2 policy evaluation, and practical implementation checklists. Section 3 applies the workflow to simulated randomized trials under scenarios with no, weak and strong heterogeneity, illustrating how often the Stage 1 gate is triggered and how much value personalization can add. Section 4 analyzes the ACTG 175 HIV trial to show how population level evidence of heterogeneity may fail to translate into exploitable individual signal. Section 5 discusses methodological limitations, external validity, regulatory considerations, and the relation of this workflow to existing guidance such as WATCH and the ICH E9(R1) estimands framework. Section 6 concludes the paper.

## 2 A conceptual framework for the hybrid workflow

Subgroup identification for treatment benefits lies at the intersection of population level inference and individual decision making. Inference addresses unobservable properties of a population, whereas decision making requires taking an action for a specific individual based on a learned rule. This structure motivates a hybrid two-stage workflow (**Figure 1**). In Stage 1, we use inference to determine whether benefit heterogeneity exists at the population level. In Stage 2, we learn and validate a decision rule that assigns treatment to individuals with appropriate error control.

### 2.1 Target estimands and identification

Let $X$ denote covariates, $A \in \{0,1\}$ a binary treatment, and $Y(0), Y(1)$ the potential outcomes under control and treatment. Our primary estimand is CATE

$$\tau(X) = \mathrm{E}[Y(1) - Y(0) \mid X]$$

We define the latent benefit label

$$Z(X) = \mathbf{1}\{\tau(X) > \delta\}$$

for a prespecified clinically meaningful margin $\delta \geq 0$. We call patients with $Z(X) = 1$ benefiters and those with $Z(X) = 0$ non-benefiters. Under a randomized trial, the assumption of consistency ($Y = Y(A)$), no interference (SUTVA), positivity ($P(A = a \mid X) > 0$ for $a \in \{0,1\}$), and randomization ($A \perp (Y(1), Y(0)) \mid X$) hold by design [10]. These conditions allow us to estimate $\tau(X)$ from observed data. In observational studies, randomization is replaced by the assumption of no unmeasured confounding, which must be justified by study design or sensitivity analyses.

Under these conditions, the latent benefit label $Z(X)$ becomes estimable through CATE estimation [11].

To make individual treatment recommendations, we use a decision rule to predict the unobserved $Z(X)$ for new patients. In this setting, a classifier outputs $\hat{Z}(X) \in \{0,1\}$ by thresholding the estimate $\hat{\tau}(X)$. Its performance can be evaluated using metrics such as the area under the uplift curve (AUQC), the expected policy value, and error control criteria that penalize treating non-benefiters. In the workflow, Stage 1 focuses on population-level hypotheses of treatment heterogeneity, providing developer whether a statistically supported benefiter subgroup exists. Stage 2 then learns and validates an individualized rule that recommends treatment for new patients. In this tutorial we consider a binary decision for a single drug. Extensions to multiple active treatments are possible but outside our scope.

## 2.2 Stage 1 of population level inference

**Formulating hypothesis tests.** We recommend formulating one of the null and alternative hypotheses to match the scientific question and the scale on which $\tau(X)$ is defined.

- **Option A: No heterogeneity**

$$H_{0A}: \tau(X) \equiv \tau_0 \qquad H_{1A}: \tau(X) \text{ varies with } X$$

This option tests whether the treatment effect is constant across the population. It provides a global screen for heterogeneity and can serve as a preliminary step before more focused tests.

- **Option B: Existence of subgroups with meaningful benefit**

$$H_{0B}: P\{\tau(X) > \delta\} = 0 \qquad H_{1B}: P\{\tau(X) > \delta\} > 0$$

This option tests whether any nonnegligible subgroup experiences a treatment effect that exceeds $\delta$. It is aligned with downstream decision making because the same margin $\delta$ defines the latent benefit label $Z(X)$.

- **Option C: Prespecified interactions.**

In a parametric model for $E[Y|A, X]$, include interaction terms $A \times X_j$ for selected $j$ and test

$$H_{0C}: \beta_{A \times X_j} = 0 \qquad H_{1C}: \beta_{A \times X_j} \neq 0$$

This option tests specific treatment-biomarker interactions. It is appropriate when there is prior biological knowledge or when the analysis plan prespecifies a small set of candidate biomarkers.

**Methods for testing heterogeneity.** We summarize several commonly used methods that test the hypotheses introduced above, demonstrating which option each method addresses and the population that the null hypothesis refers to.

- **Interaction tests in generalized linear models (target Option C; omnibus variants target Option A).** Fit a prespecified model for $E[Y|A, X]$ and test interaction coefficients with Wald or likelihood ratio tests (LRT). When the clinical concern is treatment reversal, use qualitative interaction tests such as the Gail-Simon framework and report whether the interaction is qualitative (effect varies in size but not direction) or quantitative (effect reverses direction). An omnibus joint test of all $A \times X_j$ interaction terms provide a global screen for Option A (no heterogeneity). Recent work also develops permutation interaction tests for zero-inflated biomarkers in early phase trials [12]. The population behind the null is the full trial population from which the sample is drawn.

- **GAM based smooth interactions (target Option C; omnibus variants target Option A).** Characterize treatment-biomarker interactions by fitting separate smooth functions for each treatment or by using tensor product smooths for higher dimensional interactions [13]. The resulting difference of smooths provides a nonparametric estimate of the CATE curve with pointwise confidence intervals for visualization. Evidence of interaction can be evaluated using the approximate F-test for the smooth term or with a nested model comparison. Restricted maximum likelihood (REML) estimation is recommended for selecting smoothing parameters and the basis dimension $k$ should be checked to avoid undersmoothing [14]. The null pertains to the full trial population under the specified GAM model.

- **Permutation and randomization based global tests (target Option A).** In randomized trials, permutation or Fisher randomization tests provide model-free p-values for global HTE or treatment-covariate interactions. These tests maintain nominal Type-I error under randomization and can be attractive in small samples or when model assumptions are in doubt. Examples include permutation tests for treatment-covariate interactions and randomization inference tests for unexplained treatment effect variation [15–17]. The null pertains to the randomized trial population.

- **Graphical exploration with inference bands (exploratory support for Options A and B).** The subpopulation treatment effect pattern plot (STEPP) visualizes $E[Y|A, X]$ across overlapping windows of a continuous biomarker [18]. In this method, permutation bands and bootstrap intervals are essential to avoid over interpreting random fluctuations. STEPP can provide evidence against the null hypothesis in Option A when patterns deviate from constancy. If a clinical margin $\delta$ is applied, STEPP can also apply to Option B by highlighting windows where estimated effects exceed $\delta$. It is worth noting that STEPP is exploratory and should not be the sole basis for confirmatory claims.

- **Machine learning subgroup discovery methods (exploratory support for Option B).** Estimate individualized treatment effects or benefit scores using flexible function estimators such as boosted trees and outcome regression learners, and then use ranking or thresholding of these scores to propose benefiter subgroups. They accommodate high dimensional biomarker sets and complex nonlinear interactions with variable importance measures to highlight biomarkers contributing to treatment modification. Machine learning subgroup discovery is generally exploratory due to data adaptive choice of tuning parameters, learners, and subgroup thresholds. A recent example is BioPred using A-learning and weight-learning within XGBoost to estimate individualized benefit scores, treatment rules, importance rankings, and biomarker oriented visualizations [19].

The methods above are post-hoc analytical tools applied to a completed dataset. To make a confirmatory Option B claim, a prospective design is required. We therefore include the Adaptive Signature Design (ASD) as the design counterpart that links exploratory subgroup signals to prespecified confirmatory testing.

- **Adaptive Signature Design (prospective confirmation of Option B).** ASD develops a multivariate model to predict benefits $S(X)$ in a training set and choose a threshold $c$ to define a benefiter subgroup $\mathbf{1}\{S(X) \geq c\}$ [20]. Then it performs a confirmatory test of treatment benefit within this subgroup in a held out test set. This operationalizes the Option B hypothesis of "existence of a subgroup with $\tau(x) > \delta$" using the same clinical margin $\delta$ that defines the latent benefit label $Z(X)$. Cross validated ASD variants recycle subjects across folds to improve efficiency while preserving nominal error control. The population behind the null is the trial population restricted by the prospectively defined $S(X)$. Unlike the post-hoc methods above, ASD is a trial-level design in which the modeling pipeline and decision rule are prespecified, the subgroup is defined on training set, and the benefit is prospectively tested on held-out data.

**Multiplicity and error control.** Multiplicity arises differently depending on which null hypothesis is being tested. Under Option A (no heterogeneity), if several prespecified contrasts are examined, or if an omnibus screen is followed by more focused tests, the family of tests is defined by those heterogeneity contrasts and its error must be controlled. Under Option B (existence of meaningful benefit), the family consists of multiple clinical margins $\delta$ or candidate subgroups being evaluated. Under Option C (prespecified interactions), the family is the set of all $A \times X_j$ interaction terms tested. When multiple clinical endpoints are considered (e.g., primary efficacy, key secondary efficacy, and safety), the family is expanded to include the set of endpoints in addition to the A/B/C options above.

The choice of error rate to control relies on the purpose of the analysis. For confirmatory claims such as an ASD prospective evaluation restricted to an identified subgroup, strong control of the family-wise error rate (FWER) is required. This can be achieved through gatekeeping or closed testing procedures using a prespecified trial level significance $\alpha$ [21]. In contrast, for discovery-oriented settings with many exploratory candidates, control of the false discovery rate (FDR) is often more appropriate, and weighted FDR procedures may be adopted when biology suggests prioritizing some tests over others. In either case, it is important to state explicitly the population to which the null refers, whether it is the overall trial population or a target subgroup.

Allocation of the error budget should also be prespecified. For example, one might split $\alpha$ by assigning part of it to an omnibus test for Option A at $\alpha_1$, then testing $m$ interaction terms in Option C with a Holm procedure at $\alpha_2$, and evaluating $k$ margins for Option B with a Benjamini-Hochberg (BH) adjustment at $\alpha_3$. In the case of multiple endpoints, the endpoint-level strategy (e.g., closed testing or gatekeeping) and the allocation of $\alpha$ across endpoints should be prespecified along with the clinical margins for each endpoint. Documenting these settings in advance keeps population level claims auditable.

Finally, it is critical to keep testing errors distinct between different study stages. Type-I error, FWER, and FDR are population level properties defined under repeated sampling, and they should not be confused with sample specific prediction metrics such as precision on a validation split [22]. A brief reminder belongs here, while detailed contrasts are deferred to Stage 2, where ranking and policy metrics are introduced.

With the prespecified $\alpha$ allocation across the union of Options A to C, the analysis proceeds to Stage 2 if at least one of the following criteria is met:

(i) $H_{0A}$ is rejected, indicating global heterogeneity

(ii) $H_{0B}$ is rejected for a clinically chosen $\delta$ [a]

(iii) One or more prespecified interactions in $H_{0C}$ are significant after multiplicity adjustment

If none of the criteria are satisfied or estimated $P\{\tau(X) > \delta\}$ is close to zero for all clinically relevant margins $\delta$, then the negative finding should be documented and the analysis stops. Complementary to the Stage 1 setup, recent sponsor-oriented guidance describes an exploratory workflow WATCH to plan data checks and HTE exploration before any confirmatory claims [23]. Its relation to our hybrid workflow is discussed in Section 5.

## 2.3 Stage 2 – Decision making under unobserved labels

**Estimating CATE.** Once Stage 1 establishes HTE, Stage 2 turns to individual level decisions. Because the benefit label $Z(X)$ is latent, we must first estimate the CATE under the identification conditions described in Section 2.1. Below we introduce the main families of estimation methods. Define the propensity score

$$e(X) = P(A = 1 \mid X)$$

and conditional outcome

$$\mu_a(X) = E[Y \mid A = a, X]$$

---

[a] Here, (ii) refers to a prespecified FWER-controlled Option B test for a single, clinically chosen $\delta$. When Option B is analyzed with FDR for discovery (multiple $\delta$ or candidate subgroups), it is exploratory and does not contribute to the confirmatory gate. Progression to Stage 2 is then procedural.

(i) The simplest approaches are meta-learners, which reframe CATE estimation to supervised learning problems [6].

**S-learner:** Fit a single outcome model $\hat{m}(X, A) \approx E[Y \mid A, X]$ and define

$$\hat{\tau}_S(X) = \hat{m}(1, X) - \hat{m}(0, X)$$

**T-learner:** Fit separate outcome models $\hat{\mu}_1(X)$ and $\hat{\mu}_0(X)$ and take

$$\hat{\tau}_T(X) = \hat{\mu}_1(X) - \hat{\mu}_0(X)$$

**X-learner:** Form pseudo effects $D_1 = Y - \hat{\mu}_1(X)$ for treated units and $D_0 = Y - \hat{\mu}_0(X)$ for controls, regress them on $X$ to obtain $\hat{g}_1(X)$ and $\hat{g}_0(X)$, and combine with weights $w(X)$

$$\hat{\tau}_X(X) = w(X)\hat{g}_1(X) + \{1 - w(X)\}\hat{g}_0(X)$$

where $w(X)$ is often the estimated propensity score or one of its functions.

(ii) Another type of approaches is to construct pseudo-outcomes.

**Inverse propensity weighting (IPW)** [24]:

$$Y_i^{IPW} = \frac{A_i Y_i}{e(X_i)} - \frac{(1 - A_i) Y_i}{1 - e(X_i)}$$

**Doubly robust (DR)** [25]:

$$Y_i^{DR} = \mu_1(X_i) - \mu_0(X_i) + \frac{A_i(Y_i - \mu_1(X_i))}{e(X_i)} - \frac{(1 - A_i)(Y_i - \mu_0(X_i))}{1 - e(X_i)}$$

Both $Y_i^{IPW}$ and $Y_i^{DR}$ can be regressed on $X$ to obtain an unbiased estimate of CATE, often with cross fitting to mitigate bias.

(iii) Causal forests provide a flexible and nonparametric alternative approach [7]. They directly estimate $\tau(X)$ by recursively partitioning the covariate space and averaging treatment-control contrasts within honest sample split leaves. The output $\hat{\tau}_{CF}(X)$ comes with variance estimates grounded in honesty theory, making causal forests a practical counterpart to the T-learner with built in orthogonalization.

(iv) Direct policy learning via A-learning. A-learning models the treatment contrast $\Delta(X) = E[Y \mid A = 1, X] - E[Y \mid A = 0, X]$ directly, often through a modified outcome formulation with augmentation [26,27]. Fitting a model for $\Delta(X)$ yields a treatment score $s_A(X)$. The sign of $s_A(X)$ defines an individualized rule, and $s_A(X)$ can also be used to rank patients. Unlike CATE meta-learners, A-learning targets the decision boundary and is less sensitive to misspecification of the main effects. A deep learning extension with embedded biomarker selection is DeepRAB, which uses an A-learning objective [27].

Across all these approaches, we recommend sample splitting and cross fitting to ensure nuisance functions are estimated on independent folds. This reduces bias and provides honest evaluation targets, laying the groundwork for the validation strategies described in the next subsection.

**Validating CATE estimation.** Because the true CATE is unobserved, validation in Stage 2 cannot rely on measuring accuracy against ground truth as in ordinary supervised learning. Instead, we evaluate the estimated $\hat{\tau}(X)$ by uplift curves combined with cross fitting, only using quantities identifiable from observed data to ensure unbiased evaluation. Specifically, we partition the data into $K$ folds. For each fold $k$, we estimate the propensity score $e^{(-k)}(X)$, conditional outcome $\mu_a^{(-k)}(X)$, and DR pseudo-outcome $Y_{(-k)}^{DR}$ using data from other folds. Patients in the held out fold are then ordered by their cross fitted CATE estimate $\hat{\tau}^{(-k)}(X)$, ensuring that each estimation is obtained without using the patient's own data. The cumulative uplift at the top fraction $q \in (0,1]$ of the ranked list is defined as

$$U(q) = \frac{1}{n} \sum_{i=1}^{\lfloor qn \rfloor} Y_{(i)}^{DR}$$

where $Y_{(i)}^{DR}$ are cross fitted DR pseudo-outcomes sorted by $\hat{\tau}^{(-k)}(X)$. Plotting $U(q)$ against $q$ generates the uplift curve, which shows how well the model prioritizes patients by expected treatment benefit. The area under the uplift curve (AUQC) then summarizes overall ranking performance [28]. Importantly, AUQC is a ranking metric, not a direct measure of effect modification. When the true CATE is nearly constant (i.e., little or no HTE), any well calibrated model yields an approximately linear uplift curve, and AUQC is driven mainly by the marginal treatment effect rather than by differences across patients. In such cases, a large AUQC reflects that treating most patients is beneficial but does not imply personalization.

As an alternative summary, the Area Under the Prescriptive Effect Curve (AUPEC) integrates over $q$ the difference in average outcomes between a policy that treats the top $q$ fraction and a random policy with the same budget. In randomized trials, AUPEC and its variance can be estimated without modeling assumptions, with cross-validated variants when the same data are used for learning and evaluation [29].

**Policy evaluation.** Deploying a treatment strategy in practice requires a binary decision rule. Such a rule, or policy, is equivalent to a classifier constructed from the estimated CATE and a threshold $t$:

$$\pi_t(X) = \mathbf{1}\{\hat{\tau}(X) > t\}$$

The threshold $t$ reflects a clinical tradeoff, balancing detection of benefiters against the risk of treating non-benefiters. To evaluate a candidate policy $\pi_t$, we define its value as the expected outcome in the population if that policy were deployed:

$$V(\pi_t) = \mathrm{E}[Y(\pi_t(X))]$$

We estimate $V(\pi_t)$ using the cross fitted DR estimator [30]:

$$\hat{V}_{DR}(\pi_t) = \frac{1}{n}\sum_{i=1}^{n}\left[\mu_{\pi_t(X_i)}^{(-k(i))}(X_i) + \frac{\mathbf{1}\{A_i = \pi_t(X_i)\}}{e^{(-k(i))}(X_i)}\left(Y_i - \mu_{A_i}^{(-k(i))}(X_i)\right)\right]$$

where $e^{(-k(i))}(X_i)$ and $\mu_{A_i}^{(-k(i))}(X_i)$ are propensity score and conditional outcome estimated without using subject $i$'s fold, and $\mu_{\pi_t(X_i)}^{(-k(i))}(X_i)$ denotes the predicted outcome under the action chosen by policy $\pi_t$.

$\hat{V}_{DR}(\pi_t)$ is often reported along with standard errors via the influence function or bootstrap and, when relevant, the regret of a policy, defined as the difference between the value of the optimal policy and that of $\pi_t$. In practice, the threshold $t$ can be chosen by maximizing $\hat{V}_{DR}(\pi_t)$ on held out folds, possibly subject to auxiliary constraints such as maintaining a minimum recall among predicted benefiters or ensuring fairness across subgroups.

**Safety constraint threshold selection.** In settings where harmful treatment should be rare, stakeholders may wish to control the population rate of treating patients whose treatment effect does not exceed the clinical margin $\delta$. Define the harmful event

$$\{\pi_t(X) = 1, \tau(X) \leq \delta\}$$

and fix a tolerance level $\alpha_{harm} \in (0,1)$. Because the true benefit label $Z(X)$ is latent, we implement an NP rule using cross fitted surrogates constructed from the DR pseudo-outcomes introduced above [31]. On held out folds, we treat surrogate harm: $\mathbf{1}\{Y_i^{DR} \leq \delta\}$ surrogate benefit capture: $\mathbf{1}\{Y_i^{DR} > \delta\}$ as proxies for $\tau(X_i) \leq \delta$ and $\tau(X_i) > \delta$, respectively. For each candidate threshold $t$, we then compute the harm rate among treated patients and the benefit capture rate among all patients. The NP rule selects the threshold $t$ whose harm rate does not exceed $\alpha_{harm}$ and, among all such thresholds, maximizes the benefit capture. The resulting NP-ROC band plots the tradeoff between harm and benefit capture across thresholds and highlights the chosen operating point [31].

With multiple risk endpoints, the NP rule extends to vector constraints by bounding each risk endpoint's rate. Concretely, for thresholds $t$ we compute cross fitted surrogate harms $H^{(j)}(t)$ for risk endpoints $j = 1, \ldots, J$ and choose $t$ so that $H^{(j)}(t) \leq \alpha_{harm}^{(j)}$ for all $j$, while maximizing benefit capture. The resulting NP-ROC surface summarizes the tradeoffs across thresholds under multi endpoint constraints.

## 2.4 Practical implementation

Before turning to the working example, **Table 1** summarizes common pitfalls in subgroup analysis and policy learning, together with the safeguards built into the hybrid framework. This table is intended as a practical checklist to help distinguish population level inference from individual prediction, avoid evaluation leakage, and clarify how metrics and thresholds should be interpreted. In practice, implementation reduces to concrete software choices, pragmatic default tuning rules, and transparent reporting of key analysis decisions. **Table 2** provides an implementation template for these elements and can serve as a checklist when applying the workflow in new trials. The simulation study in Section 3 and the real data analysis in Section 4 follow this template. Deviations from the defaults in **Table 2** are noted explicitly where they occur.

# 3 The hybrid workflow in simulated trials

To illustrate the two-stage framework, we conduct a simulation study under three mechanisms of data generation: no, weak, and strong HTE. In each scenario, population level inference and individual decision making are combined into a coherent workflow. The simulations highlight how the hybrid approach separates exploratory inference from individualized decision rules, how often the Stage 1 gate is triggered, and how much incremental value can be obtained from personalization under different degrees of HTE.

## 3.1 Data generation

We simulate randomized trials with $n$ patients and baseline covariates $X = (X_1, X_2, X_3)$. Treatment $A \in \{0,1\}$ is assigned at ratio 1:1. Outcomes $Y \in \{0,1\}$ follow a logistic model with baseline log odds

$$\eta_0(X) = \beta_0 + \beta^T X$$

and treatment log odds increment

$$\tau(X) = \gamma_0 + \gamma_1 X_1$$

so that the biomarker $X_1$ is the primary effect modifier. The conditional CATE is

$$\tau(X) = P(Y = 1 \mid A = 1, X) - P(Y = 1 \mid A = 0, X)$$

We fix a clinical margin to $\delta$ and define the latent benefit label

$$Z(X) = 1\{\tau(X) > \delta\}$$

which encodes whether a patient would benefit sufficiently from treatment. All three scenarios share the same baseline model $\eta_0$ and covariate distribution. They differ in the parameters $\gamma_0$ and $\gamma_1$ that define the treatment effect:

(i) **No HTE.** We set $\gamma_1 = 0$, so that $\tau(X) \equiv \gamma_0$ is constant. This represents a trial with a nonzero average treatment effect but no true effect modification by $X_1$. Any apparent heterogeneity arises from sampling variability.

(ii) **Weak HTE.** We choose $\gamma_1$ small and positive, so that $\tau(X)$ increases with $X_1$ but remains positive for most patients. Only a modest fraction of patients exceeds the clinical margin $\delta$, and the contrast between benefiters and non-benefiters is mild.

(iii) **Strong HTE.** We choose $\gamma_0 < 0$ and $\gamma_1 > 0$ large enough that treatment is harmful for patients with low $X_1$ but beneficial among those with high $X_1$. The average treatment effect can be slightly negative, while a sizable subgroup with large $X_1$ enjoys clinically meaningful benefit.

For each scenario, we generate 200 independent trial replicates. **Table 3** provides the full details of the simulation parameters.

### 3.2 Stage 1: inference of treatment heterogeneity

Stage 1 tests for treatment heterogeneity using two of the prespecified options from Section 2.2.

- **Option A (global heterogeneity).** We fit a logistic model for $Y$ with main effects in $X$ and compare it to a model that additionally includes all $A \times X$ interactions. An omnibus likelihood ratio test (LRT) is used to test the null hypothesis of no interaction between treatment and covariates.

- **Option C (prespecified interactions).** We fit a logistic model that includes prespecified interactions $A \times X_1, A \times X_2, A \times X_3$. Wald tests for individual interaction terms are performed and adjusted for multiplicity using Holm's method.

In each simulated trial, we proceed to Stage 2 if either the omnibus LRT or at least one prespecified interaction remains significant at $\alpha = 0.05$ after multiplicity adjustment. Otherwise, the workflow stops after Stage 1.

Across the 200 replicates, the proceed rate in **Table 3** summarizes how often the gate is triggered under each scenario. Under No HTE, the gate is triggered in only 8.5% of replicates, reflecting a moderate but controlled type-I error rate when there is truly no effect modification. Under Weak

HTE, the proceed rate rises to 65% of replicates, indicating that the Stage 1 tests have reasonable power but do not declare heterogeneity in every replicate. Under Strong HTE, the proceed rate is 100%, showing the workflow always progresses to Stage 2 when strong qualitative interaction is present. **Figure 2** visualizes one representative replicate under the Strong HTE. The STEPP in panel A shows that windowed risk differences are negative at low $X_1$ and strongly positive at high $X_1$, reinforcing the heterogeneity evidence from Options A and C. **Supplementary Figures S1** and **S2** show representative replicates under the Weak and No HTE, respectively.

### 3.3 Stage 2: CATE estimation, validation, and policy value

In trial replicates where the Stage 1 gate is triggered, we proceed to individualized decision making.

**CATE estimation.** We estimate the CATE $\tau(X)$ using a causal forest fitted to the full set of baseline covariates. Cross fitting is used for estimating propensity scores and outcome regressions. The causal forest is trained on one subset of folds and evaluated on held out folds to avoid evaluation leakage and to obtain DR pseudo-outcomes for validation.

**Ranking performance.** Within each replicate, we order patients by their estimated CATE $\hat{\tau}(X_i)$, accumulate the cross fitted DR pseudo-outcomes to form an uplift curve, and compute the AUQC. AUQC increases when the marginal treatment effect is larger and treatment benefit is more concentrated among top ranked patients. Because the average treatment effect is positive in all three scenarios, AUQC is nonzero even under No HTE. Moreover, AUQC is driven partly by the overall treatment effect. In our parameterization, the marginal effect is smallest under Weak THE. Therefore, its uplift curve is compressed and the mean AUQC (9.9) is much lower than under No HTE (85.0) and Strong HTE (86.1), even though only Strong HTE exhibits pronounced effect modification (**Table 3**). These numbers are not pure measures of heterogeneity but reflect the total amount of treatment benefit in each scenario and how that benefit is distributed across risk strata. For this reason, we view AUQC as a descriptive ranking metric, whereas the value gain below isolates the benefit of personalization beyond the non-personalized policy.

**Policy threshold and value gain.** For any candidate threshold $t$, the individualized treatment rule is

$$\pi_t(X) = \mathbf{1}\{\hat{\tau}(X) > t\}$$

which treats patients whose estimated CATE exceeds $t$. We estimate the value of this policy using the cross fitted DR estimator described in Section 2.3, yielding $\hat{V}(t)$ on the outcome. Within each replicate, we select the policy threshold

$$t^* = \mathrm{argmax}_t \hat{V}(t)$$

over a grid of thresholds and compute the corresponding policy value $t^*$. To quantify the benefit of personalization, we compare $\hat{V}(t^*)$ with the policy of treating all or treating none. Specifically, we define the value gain

$$\Delta V = \hat{V}(t^*) - \max\{\hat{V}(\text{treat all}), \hat{V}(\text{treat none})\}$$

which measures the improvement obtained by applying the learned individualized rule instead of the fixed treatment strategy.

**Table 3** reports the average value gain $\Delta V$ across replicates for each scenario. Under No HTE, $\Delta V$ is essentially zero, even in replicates where Stage 1 spuriously suggests heterogeneity. Weak HTE produces a modest average gain of 0.017, corresponding to an improvement of about 1-2% in outcome probability relative to the fixed treatment policy. Under Strong HTE, the gain rises to approximately 0.056, indicating that an individualized rule can increase the probability of a favorable outcome by more than 5% compared with treating all or treating none. **Figure 2** illustrates a typical replicate of Strong HTE, in which the policy value curve has a clear interior maximum and the optimal cutoff $t^*$ lies well above the prespecified clinical margin $\delta$, confirming the benefits of the individualized rule learned in Stage 2. The analogous visualizations for Weak and No ETH scenarios are shown in **Supplementary Figures S1** and **S2**.

### 3.4 Safety constrained threshold

Finally, we examine the NP safety rule from Section 2.3 that prioritizes limiting harm over maximizing value. For each replicate and candidate threshold $t$, we compute cross fitted DR for the harm rate and benefit capture rate. We fix a tolerance of harm rate $\kappa = 0.10$ and apply the NP rule on held out folds. Among all thresholds whose estimated harm rate does not exceed $\kappa$ based on one-sided Wilson upper confidence bounds, we choose the one that maximizes benefit capture. If no threshold satisfies the harm constraint, we report the best attainable operating point that minimizes harm subject to achieving nontrivial benefit capture.

Across all three scenarios and their replicates, the 10% harm constraint is stringent relative to the available signal. No threshold satisfies the NP bound, and the rule always returns a best attainable point rather than a fully feasible one. **Figure 2D, Supplementary Figures 1D** and **2D** display the average harm-benefit frontiers for the typical replicates of three scenarios. In the No HTE and Weak HTE scenarios, the frontier is relatively close to a vertical line, indicating that a large reduction in harm can be achieved only at the cost of sacrificing nearly all benefit. In the Strong HTE scenarios, the frontier shifts slightly toward higher benefit capture, but the best attainable point still corresponds to a harm rate around 40% with modest benefit capture.

## 4 The workflow in a real trial ACTG 175

We now apply the two-stage workflow to a publicly available randomized trial to illustrate how population level inference and individual decision making interact in practice. Consistent with the workflow, Stage 1 establishes whether credible THE exists with multiplicity control. If such evidence exists, we proceed to Stage 2, where we learn and validate an individualized rule using cross fitted and DR metrics with an NP harm constraint. All evaluations in Stage 2 are cross fitted to avoid leakage.

### 4.1 Dataset and preprocessing

ACTG 175 is a randomized, multicenter HIV clinical trial that compared zidovudine (AZT) with didanosine (ddI) and combination regimens that paired AZT with ddI or zalcitabine (ddC) [32]. In our analysis, we set AZT as $A = 0$ and combination therapy (AZT+ddI or AZT+ddC) as $A = 1$, a binary contrast aligned with clinical use. We define a binary outcome $Y = 1$ if no event is observed by 96 weeks and $Y = 0$ otherwise. Candidate baseline covariates $X$ include demographics, disease status, and laboratory measures, with post randomization variables excluded to avoid leakage. We remove observations with missing values on $(A, Y, X)$. This preprocessing results in a clean dataset with sample size $N = 1578$ and covariate number $p = 16$.

### 4.2 Stage 1: Population level inference

We test two prespecified hypotheses defined in Section 2.2.

**Option A (global heterogeneity)**. We compare a logistic model of main effects to a model that adds all $A \times X$ interactions via an LRT test. The omnibus LRT rejects no heterogeneity ($\chi^2 = 37.2, df = 16, p = 0.002$), providing population level evidence that treatment effects vary with baseline covariates.

**Option C (prespecified interactions).** We prespecify baseline Karnofsky and CD4 as two biologically interpretable moderators and perform Wald tests within the interaction model. The Karnofsky × treatment term remains significant after Holm adjustment (adjusted $p = 0.015$), whereas the CD4 × treatment term does not. Together with the STEPP plots in **Figures 3A and 3B**, these results justify proceeding to Stage 2 under our gate rule.

### 4.3 Stage 2: individual learning and validation

All quantities in Stage 2 are computed out of fold, and therefore, every prediction used for evaluation comes from a model that was not trained on that individual. These metrics target decision making and are distinct from the Stage 1 testing error described in Section 2.2. The CATE is estimated by a causal forest model.

**Ranking performance.** We order patients by the causal forest estimate $\hat{\tau}(X)$, accumulate the cross fitted DR pseudo-outcomes to obtain the uplift curve $U(q)$, and then compute AUQC. In the ACTG 175 dataset, the AUQC is approximately 0.068 (**Figure 3C**). It lies close to the random ranking baseline, which is roughly one half of $U(1)$, the estimated average treatment effect (ATE) at $q = 1$. Interpreted in units of incremental outcome, this indicates that, given these baseline features and this endpoint, the learned score ranks benefiters only slightly better than chance.

**Decision quality.** For the class of threshold policies $\pi_t(X) = \mathbf{1}\{\hat{\tau}(X) \geq t\}$, we estimate the policy value $V(\pi_t)$ using a cross fitted DR estimator. The resulting value curve attains its maximum at the smallest threshold considered and then decreases monotonically as $t$ increases, implying that "treat nearly all" dominates within this family of rules (**Figure 3D**). This pattern is consistent with a positive overall ATE together with near random ranking.

**Safety constrained selection.** With clinical margin $\delta = 0$, we define surrogate harm and benefit indicators from DR outcomes and, for each threshold $t$, evaluate the harm rate among treated patients $P(\text{harm} \mid \pi_t(X) = 1)$ and the benefit capture rate among treated patients $P(\text{benefit} \mid \pi_t(X) = 1)$. Under a harm tolerance $\alpha_{harm} = 0.10$, no threshold satisfies the constraint, as the estimated harm rate remains approximately 0.32-0.34 across thresholds (**Figure 3E**). It reflects that the selection is close to random and that the population fraction of non-benefiters is well above 10%. In this setting we therefore report the best attainable frontier on the NP-ROC curve.

## 4.4 Discussion of the real data results

This case study illustrates a key message of the workflow. Stage 1 provides population level evidence that HTE exists and that Karnofsky is a candidate predictive biomarker in the inferential sense. Stage 2, however, shows that with the measured baseline covariates and this 96-week endpoint, the heterogeneity is not predictably learnable for individualized decisions. The uplift is near the random baseline, the best policy within threshold rules is effectively treating all patients, and a stringent NP harm bound is infeasible. In our framework, population level heterogeneity is necessary but not sufficient for actionable personalization. We require evidence of Stage 2 that an individualized rule improves expected outcomes and meets safety tolerances before recommending it. Absent such evidence here, a uniform strategy is preferable for this contrast and endpoint, while Karnofsky remains a hypothesis generating signal for future work with additional biomarkers, an alternative prespecified endpoint, or prospective validation.

## 5 Interpretation, limitations, and regulatory implications

While the hybrid framework provides a structured route from HTE detection to individualized policies, its guarantees rely on several modelling, design, and governance choices. We summarize key methodological limitations of the workflow, issues of external validity and fairness, and regulatory and operational considerations for deployment. We then position the framework relative to WATCH and other HTE workflows and discuss the choice of the clinical margin δ and its link to estimands.

## 5.1 Methodological limitations

First, the analysis in Stage 2 relies on standard identification conditions such as consistency, no interference, ignorability, and positivity, as outlined in Section 2.1. When estimated propensity scores are close to 0 or 1, both the DR pseudo-outcomes and the DR policy value estimator can become unstable. Cross fitting reduces overfitting bias but cannot restore information that is fundamentally absent in this scenario [33]. These issues are visible in the simulations, where value curves are flat in the No HTE scenario and noisy in Weak HTE, and in ACTG 175, where the CATE score cannot rank patients much better than chance despite significant Stage 1 interactions.

Second, the NP harm constraint may be conservative or infeasible in realistic sample sizes. In our simulations and ACTG 175 analysis, no threshold satisfies a 10% harm tolerance, so the NP rule returns a best attainable operating point rather than a fully feasible one. This should be interpreted as a property of the data and constraint, not as a failure of the learning algorithm. If the true fraction of non-benefiters among treated patients exceeds the tolerance, no threshold rule will satisfy the bound.

Third, the CATE estimation depends on nuisance models and tuning parameters. Even with cross fitting, variance can be high when the signal is weak, leading to noisy uplift and value curves. To prevent over interpretation, we suggest reporting uncertainty intervals for AUQC and policy values, as well as emphasizing value gain over non-personalized strategies rather than absolute performance. When bootstrap uncertainty is desired but computational cost is prohibitive, scalable bootstrap variants have been proposed for large datasets [34].

Finally, threshold selection introduces another source of optimism if it is tuned post hoc on the same data used for estimation [35]. The workflow mitigates this through cross fitted evaluation and by defining threshold choice in terms of prespecified rules (e.g., maximizing DR value over a grid, or applying an NP harm constraint). Nevertheless, analysis plans should document the policy family and selection rule in advance to keep Stage 2 decisions auditable.

## 5.2 External validity and transportability

Policies learned from a single trial may not transport unchanged to other settings. If the distribution of baseline covariates differs between the study and the target population, estimates of AUQC and policy value can be biased unless reweighted to the target covariate distribution [36]. Importance weighting of the cross fitted DR estimators can partially adjust for such differences, but they require reliable external information on the target population and remain sensitive to extreme weights.

A more difficult challenge is mechanism shift, where the conditional outcome model changes because of evolving standard of care or patient management. Under such shifts, a CATE model trained in the original trial may rank individuals incorrectly even if the marginal covariate distribution is similar, making retraining or formal transportability analyses necessary. When retraining is infeasible, deployment should be restricted to covariate regions where strong trial evidence exists.

External validity also interacts with equity. A policy that performs well on average may still concentrate harm in certain subgroups [37]. Applying the NP rule within subgroups and reporting subgroup specific harm and benefit capture rates can make such tradeoffs explicit. If subgroup specific constraints are infeasible, the analysis should state clearly which subpopulations are insufficiently supported by the data and where further evidence or new studies are required.

### 5.3 Regulatory and governance considerations

From a regulatory perspective, transparency and prospective planning are as important as statistical validity. For confirmatory trials, the analysis plan should prespecify the Stage 1 hypotheses and multiplicity strategy, the learner family to be used in Stage 2, the cross fitting scheme, the primary policy evaluation metrics, and the rule for choosing the operating threshold $\tau$ [38]. Regulatory agencies are increasingly attentive to estimands, thus the chosen estimand, the clinical margin $\delta$, and the target deployment population should be aligned with ICH E9(R1) and described in the protocol.

When deployment depends on the learned policy, regulators may also expect prospective or external confirmation of policy performance. Beyond the initial analysis, governance is required once a policy is in use, such as monitoring outcome rates and harm rates over time, checking for covariate and performance drift, and updating or suspending the policy when drift is detected [39]. These requirements are consistent with the emerging guidance of machine learning operations in health care, which views model maintenance as an ongoing process rather than a one-time deliverable. Related work already positions machine learning models as supplementary evidence in drug safety assessment and explicitly emphasize validation and uncertainty quantification [40,41].

Fairness and human oversight are additional regulatory considerations. Even when overall harm is controlled by an NP rule, sponsors should avoid situations where demographic or clinical subgroups bear disproportionate risk. Subgroup specific constraints, clinician override mechanisms, and clear uncertainty summaries help keep the policy in the realm of decision support rather than automated allocation. The secure handling of individual data for cross fitted validation and estimation is also a prerequisite for regulatory acceptance [42].

## 5.4 Relation to WATCH and other HTE workflows

Recent sponsor-oriented guidance has proposed WATCH as a workflow to assess HTE in drug development [23]. WATCH is designed as an exploratory framework for completed randomized trials, helping sponsors plan data checks, select analysis methods, and summarize the credibility of HTE findings before any confirmatory claims are made. WATCH and our workflow occupy adjacent parts of the HTE pipeline. WATCH is primarily exploratory and sponsor facing. It helps characterize where HTE signals may lie, which biomarkers merit closer attention, and how robust those signals appear under alternative analyses. Our workflow is primarily inferential and focuses on decision. It requires prespecified hypotheses and error control at Stage 1. Conditional on passing that gate, it delivers an auditable individualized policy with explicit estimates of value, harm, and benefit capture. To our knowledge, WATCH does not prescribe cross fitted policy evaluation or NP safety rules, whereas these are core components of Stage 2.

In practice, the two workflows can be combined. In early development or in external datasets, a WATCH exploratory assessment can be used to screen biomarkers, understand data limitations, and refine scientific questions. Insights from this exploratory work can then inform the prespecified Stage 1 hypotheses and analysis plan when a pivotal trial is designed. After the trial reads out, the hybrid workflow can be applied to that prespecified plan. Stage 1 provides confirmatory evidence for or against clinically meaningful HTE, and Stage 2 determines whether an individualized rule improves outcomes to justify deployment.

Our workflow adds an explicit gate at population level and an NP safety layer compared to existing tutorials, which typically start from a given trial and proceed directly to Stage 2 estimation and value evaluation. Our workflow clarifies that modern machine learning tools for HTE are not alternatives to traditional subgroup inference but rather can be embedded downstream of carefully controlled Stage 1 analyses to provide individualized recommendations that remain compatible with regulatory expectations for confirmatory trials.

## 5.5 Choice of the clinical margin $\delta$ and estimands

Throughout the workflow, the clinical margin $\delta$ plays a central but conceptually distinct role from the operating threshold used in Stage 2. From an estimand perspective, $\delta$ is part of the target rather than the estimation procedure. Current regulatory guidance encourages specifying effect measures and clinically meaningful differences at the design stage, including the estimand population, endpoint, and summary measure [38]. In our framework, the estimand is typically a marginal or conditional risk difference, odds ratio, or survival contrast defined in Section 2.1. $\delta$ encodes how large a contrast must be at the patient level for treatment to be considered clinically worthwhile, considering toxicity, burden, and alternatives. A natural choice is to place $\delta$ on the same scale as the primary efficacy estimand guided by historical evidence and clinical input. The exact value will vary by indication, but it should be fixed *a priori* and justified in the protocol or analysis plan rather than tuned to the observed data.

It is important to distinguish this design margin $\delta$ from the operating threshold $\tau$ used in Stage 2. The label $Z(X)$ and the NP harm event $\{\tau(X) < \delta\}$ are defined in terms of the true and unknown CATE. In contrast, the Stage 2 policy $\{\hat{\tau}(X) \geq \tau\}$ is defined by thresholding a learned score $\hat{\tau}(X)$. Here $\tau$ is an algorithmic tuning parameter chosen to optimize a value criterion or satisfy a harm constraint on cross fitted DR estimates. As noted in **Table 1**, equating $\delta$ and $\tau$ would conflate these roles. It would hardwire the design margin into the prediction rule, ignore the empirical value curve, and risk poor performance if the learned score $\hat{\tau}(X)$ is biased or noisy. In our workflow, $\delta$ remains fixed as part of the estimand definition, whereas $\tau$ is allowed to adapt to the data under a leakage safe evaluation scheme.

## 6 Conclusion

We presented a hybrid two-stage workflow that links population level inference on HTE with individual policy learning. Stage 1 provides auditable evidence for or against clinically meaningful heterogeneity. Conditional on passing this gate, Stage 2 learns and evaluates individualized treatment rules using cross fitted DR metrics and NP harm constraints. The simulation and ACTG 175 case study illustrate how this framework separates exploratory subgroup signals from actionable personalization, clarifying when a uniform versus individualized strategy is preferable. Taken together, the workflow offers a concise, regulator compatible blueprint for moving from subgroup findings to transparent and clinically interpretable treatment policies

## Conflicts of Interest

## Code and Data Availability

All code needed to reproduce the results of this study are openly accessible at GitHub repository https://github.com/xnnba1984/A-Tutorial-on-the-Hybrid-Two-Stage-Workflow.

The ACTG 175 clinical trial data used in this study is extracted from the R package speff2trial.

## Funding

Author Lin Wang is supported by the National Science Foundation (DMS-2413741) and the Central Indiana Corporate Partnership AnalytiXIN Initiative.

## Author Contributions

Nan Miles Xi implemented the statistical and computational methods, conducted the simulation and real data analysis, and drafted the manuscript. Xin Huang contributed to methodological refinement, interpretation of results, and manuscript revision. Lin Wang led the conceptualization and overall study design, supervised the methodological development and analysis, and provided critical review and editing of the manuscript. All authors reviewed and approved the final version of the manuscript.

# Figures

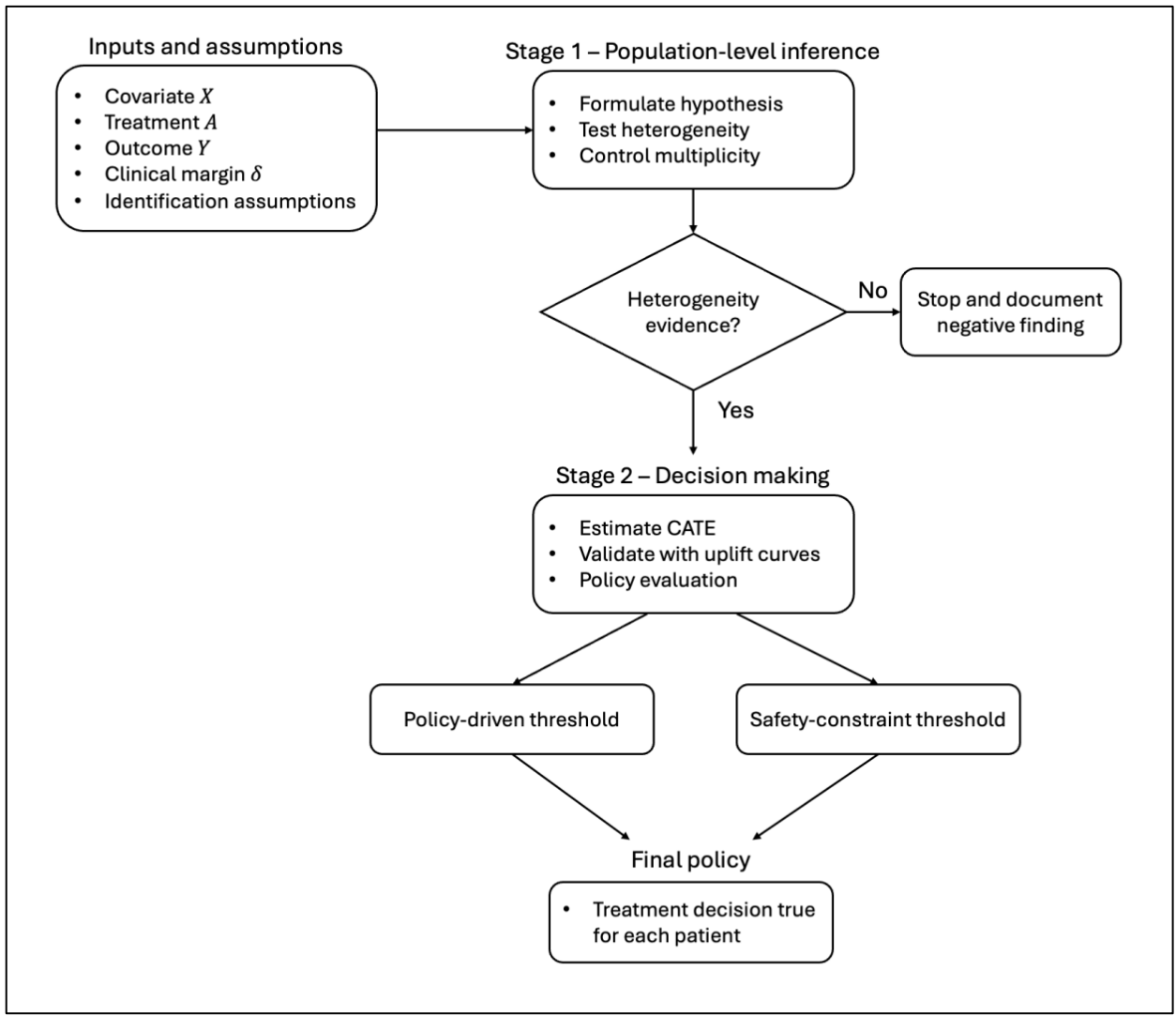

**Figure 1. Hybrid two-stage workflow.** Stage 1 tests for HTE at the population level with multiplicity control, stopping if no evidence is found. If heterogeneity is detected, Stage 2 estimates and validates CATE, evaluates policies, and selects either a policy driven or safety constrained threshold. The final output is an individualized treatment policy with explicit error control.

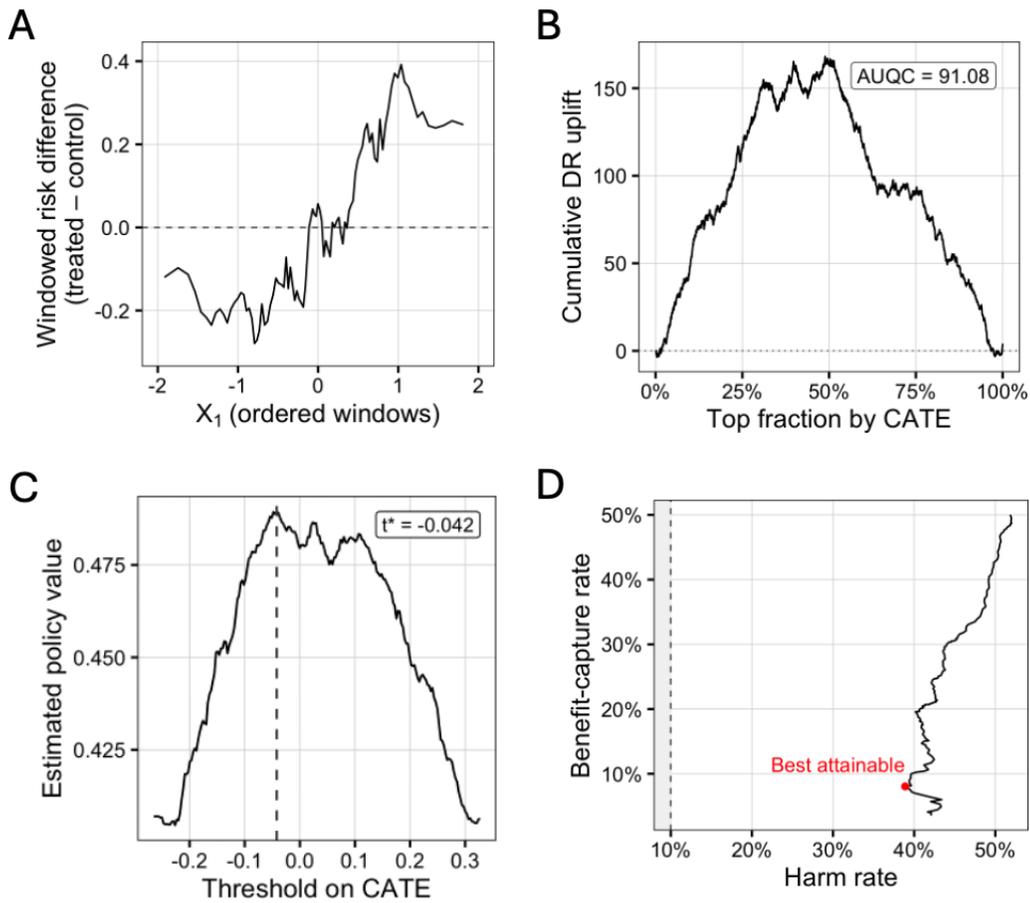

**Figure 2. Results from one Strong HTE replicate applying the two-stage workflow.** (A) STEPP exploration of treatment effect heterogeneity along biomarker $X_1$. The windowed risk difference is negative at low $X_1$ and transitions to positive values at high $X_1$, providing visual evidence of effect modification. (B) Uplift curve based on cross fitted DR pseudo-outcomes, showing concentration of treatment benefit among top ranked patients by estimated CATE. The area under the uplift curve (AUQC) is 91.08. (C) Cross fitted DR policy value $\hat{V}(\pi_t)$ as a function of the CATE threshold $t$. The curve attains an interior maximum at $t^* = -0.042$, defining the optimal threshold for this simulated trial. (D) Neyman-Pearson (NP) rule summarizing harm-benefit tradeoffs across thresholds. No threshold satisfies the 10% harm constraint. The best attainable point achieves a harm rate of 0.389 with a benefit capture rate of 0.081.

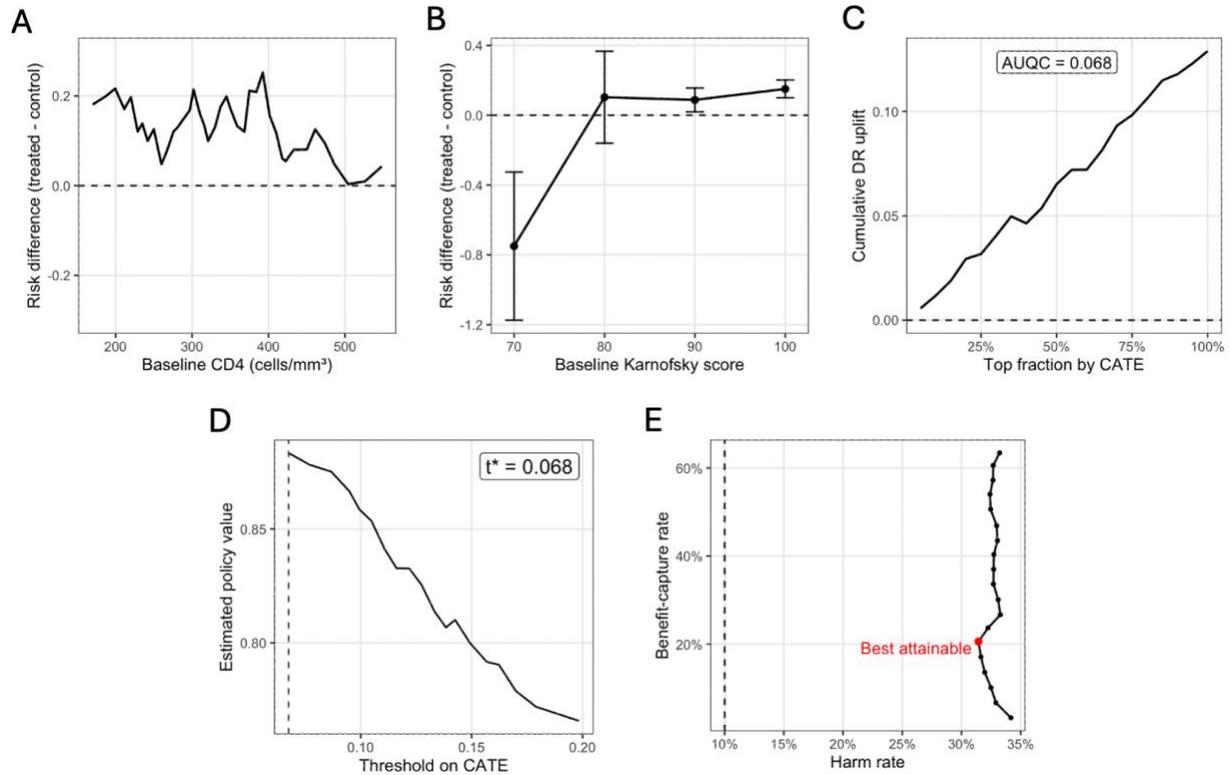

**Figure 3. ACTG 175 clinical trial applying the two-stage workflow. (A)** STEPP exploration of HTE along baseline CD4. The windowed risk difference fluctuates around zero without a clear monotone pattern, suggesting limited evidence that CD4 alone is a strong modifier of treatment effect. **(B)** Karnofsky specific risk differences. Patients with poor performance status appear to be worse on combination therapy, whereas those with higher Karnofsky scores show modest positive risk differences. **(C)** Uplift curve based on cross fitted DR pseudo-outcomes, with AUQC = 0.068. The nearly linear shape and small AUQC show that the learned CATE is unable to prioritize patients by treatment benefit beyond random ordering. **(D)** Cross-fitted DR policy value as a function of the threshold on the CATE. The curve is maximized at the smallest threshold $t^* = 0.068$, indicating that an individualized threshold rule does not improve upon a treat all policy. **(E)** NP harm-benefit frontier computed from DR surrogates. The best attainable point minimizes the estimated harm rate at 33% while capturing about 20% of potential benefit, reinforcing that exploitable heterogeneity is not present for this outcome in ACTG 175.

# Tables

**Table 1. Common pitfalls in subgroup analysis and policy learning, as well as remedies provided by the hybrid framework.**

| Pitfall | Remedy |
|---|---|
| Treating no rejection at Stage 1 as proof of the null, or confusing validation precision with FDR. | Keep Stage 1 testing errors (power, FWER, FDR) clearly separate from Stage 2 prediction metrics (AUQC, $\hat{V}_{DR}(\pi_t)$). Report confidence intervals or power for Stage 1 claims. Never label prediction performance as FDR. |
| Using the clinical margin $\delta$ as the operating cut point or letting it change with data. | Fix $\delta$ a priori to define benefit. Choose the operating threshold $t$ adaptively by maximizing policy value or using the NP rule. |
| Treating predictions from outcome model as ground truth and computing accuracy/ROC. | Validate with uplift metrics such as AUQC and cross fitted DR policy value estimates. Do not convert surrogates into true labels. |
| Reusing the same folds to fit nuisance functions, tune thresholds, and evaluate performance. | Use cross fitting for nuisance estimation and reserve a strict evaluation split. If tuning $t$, use a separate test split or prespecified cross validated rule. |
| Interpreting STEPP visualizations as confirmatory discoveries. | Treat STEPP as exploratory. Confirmatory claims must come from prespecified Stage 1 tests (options A-C). |
| Reading AUQC as a probability or calling it "area under ROC." | AUQC is measured in incremental outcome units, not restricted to [0,1]. Label axes clearly and provide sample size context. |
| Deploying a learned policy without considering transportability. | State the intended deployment population explicitly. If it differs from the trial, add assumptions for transportability or plan external validation. |

**Table 2. Practical implementation summary for the hybrid two-stage workflow**

| Category | Task | Recommended implementation |
|---|---|---|
| **Software (R language)** | Data preprocessing | `data.table` or `dplyr` for preprocessing. `glm` and `survival` for generalized linear and Cox models. |
| | Stage 1 GLM tests | Fit prespecified interaction models using `glm`. Obtain Wald or likelihood ratio tests for Options A/C. |
| | Stage 1 GAM interactions | Use `mgcv::gam` with treatment-covariate smooths and REML smoothing. Compare models with and without interaction smooths. |
| | STEPP visualizations | `stepp` for subpopulation treatment effect pattern plots with permutation or bootstrap bands. |
| | Stage 2 CATE estimation | Causal forests via `grf`. Alternatives include meta-learners or A-learning implemented with tree based or boosting methods. |
| | DR and policy evaluation | Implement cross fitting and DR pseudo-outcomes by combining propensity and outcome models with the formulas in Section 2.3. Reuse the same infrastructure for uplift curves, policy value curves and NP quantities. |
| | NP rule | Construct surrogate harm and benefit capture rates from cross fitted DR pseudo-outcomes and compute the NP-ROC frontier via custom code. |
| **Default settings** | Cross fitting design | Use $K = 5$ folds as a default. $K = 2$ may be used in small samples. Construct folds stratified by treatment, and use the same partition for nuisance estimation, CATE learning and evaluation. |
| | CATE learner tuning | For causal forests, start with package defaults and increase tree count if value or uplift curves are unstable. For meta-learners/A-learning, tune the underlying regressors by cross validation within the training folds only. |
| | Policy thresholds | Evaluate policies on a grid of quantiles of the estimated CATE, plus treat all and treat none thresholds. |
| | Harm tolerance for NP rule | Choose the harm tolerance at the design stage (5-10%) and keep it fixed across analyses of the same endpoint. |
| | Uncertainty quantification | For AUQC and policy values, use influence function standard errors or subject level nonparametric bootstrap. For NP constraints, accompany point estimates of harm with one-sided confidence bounds by Wilson or bootstrap. |
| **Reporting** | Data and estimand | Report definitions of $(X, A, Y)$, eligibility criteria, missing data handling, primary estimand, and clinical margin $\delta$ used for the latent benefit label. |
| | Stage 1 design | State which of options A-C were tested, prespecified biomarkers and STEPP, the overall $\alpha$, multiplicity strategy and the gate rule for proceeding to Stage 2. |
| | Stage 2 learning and evaluation | Describe the chosen CATE learner, tuning parameters, and cross fitting scheme. Explain how uplift curves, AUQC, and policy values were computed and how uncertainty was obtained. |
| | Policy selection and safety | Specify the policy family, the grid of thresholds considered, the rule used to select the operating threshold, the harm tolerance, and definitions of harm and benefit capture. |
| | Summary of findings | Summarize whether the Stage 1 gate was triggered, the recommended policy in Stage 2, its estimated value, value gain over non-personalized strategies, and the associated harm and benefit capture rates. |

**Table 3. Summary of the simulation parameters and results from the two-stage workflow applied to simulated trials.**

| Process | Component | Description/Result |
|---|---|---|
| Simulation parameters | Sample size | $n = 2000$ patients per trial replicate |
| | Baseline covariates | $X = (X_1, X_2, X_3)$<br>$X_1, X_2 \sim N(0,1); X_3 \sim Bin(0.5)$ |
| | Treatment assignment | $A \in \{0,1\}$, randomized 1:1 |
| | Baseline log odds | $\eta_0(X) = -0.6 + 0.6X_1 - 0.2X_2 + 0.3X_3$ |
| | Treatment log odds increment | No ETH: 0.4<br>Weak ETH: $-0.05 + 0.3X_1$<br>Strong ETH: $-0.05 + 1.0X_1$ |
| | Clinical margin | $\delta = 0.03$ |
| Stage 1 | Hypothesis | Option A: omnibus LRT<br>Option C: Wald tests for $A \times X_j$ (Holm adjusted) |
| | Proceed rate to Stage 2 | No HTE: 8.5%<br>Weak HTE: 65%<br>Strong HTE: 100% |
| Stage 2 | CATE estimation | Causal forest, cross-fitted propensity scores, outcome regressions, and DR pseudo-outcomes |
| | Ranking performance (mean AUQC) | No HTE: 85.0<br>Weak HTE: 9.87<br>Strong HTE: 86.1 |
| | Policy value gain (average) | No HTE: 0.002<br>Weak HTE: 0.017<br>Strong HTE: 0.056 |
| | NP safety rule | For all three scenarios, no threshold satisfies the 10% harm constraint |

**Supplementary**

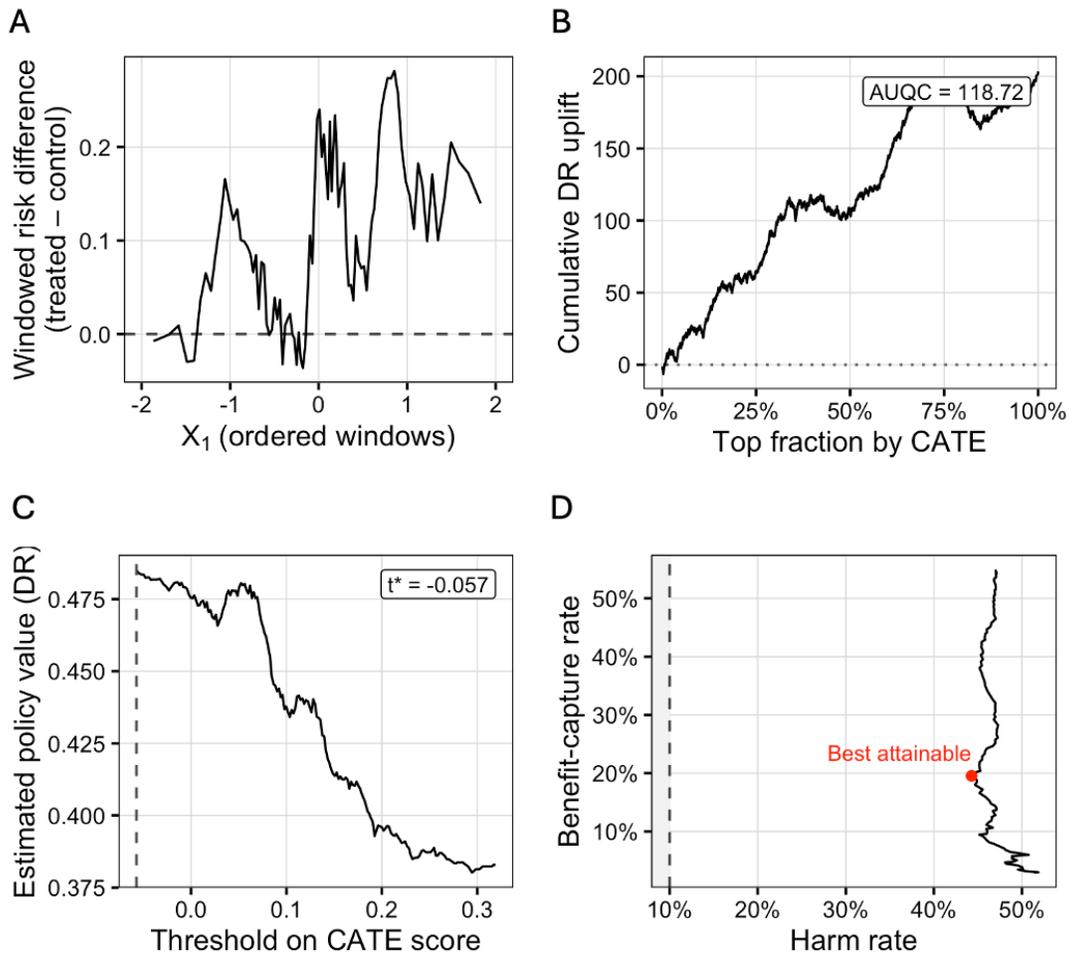

**Supplementary Figure S1. Results from one No HTE replicate applying the two-stage workflow.** (A) STEPP exploration of treatment effect heterogeneity along biomarker $X_1$. The curve fluctuates around zero and any upward trend at high $X_1$ is due to sampling variability. (B) The AUQC is measured in incremental outcome units and is positive because the average treatment effect is beneficial for most patients. (C) Cross fitted DR policy value $\hat{V}(\pi_t)$ as a function of the CATE threshold $t$. The vertical dashed line marks the optimal threshold $t^* = -0.057$, indicating that the best rule treats all patients and that personalization yields no value gain. (D) NP rule summarizing harm-benefit tradeoffs across thresholds. No threshold satisfies the 10% harm constraint.

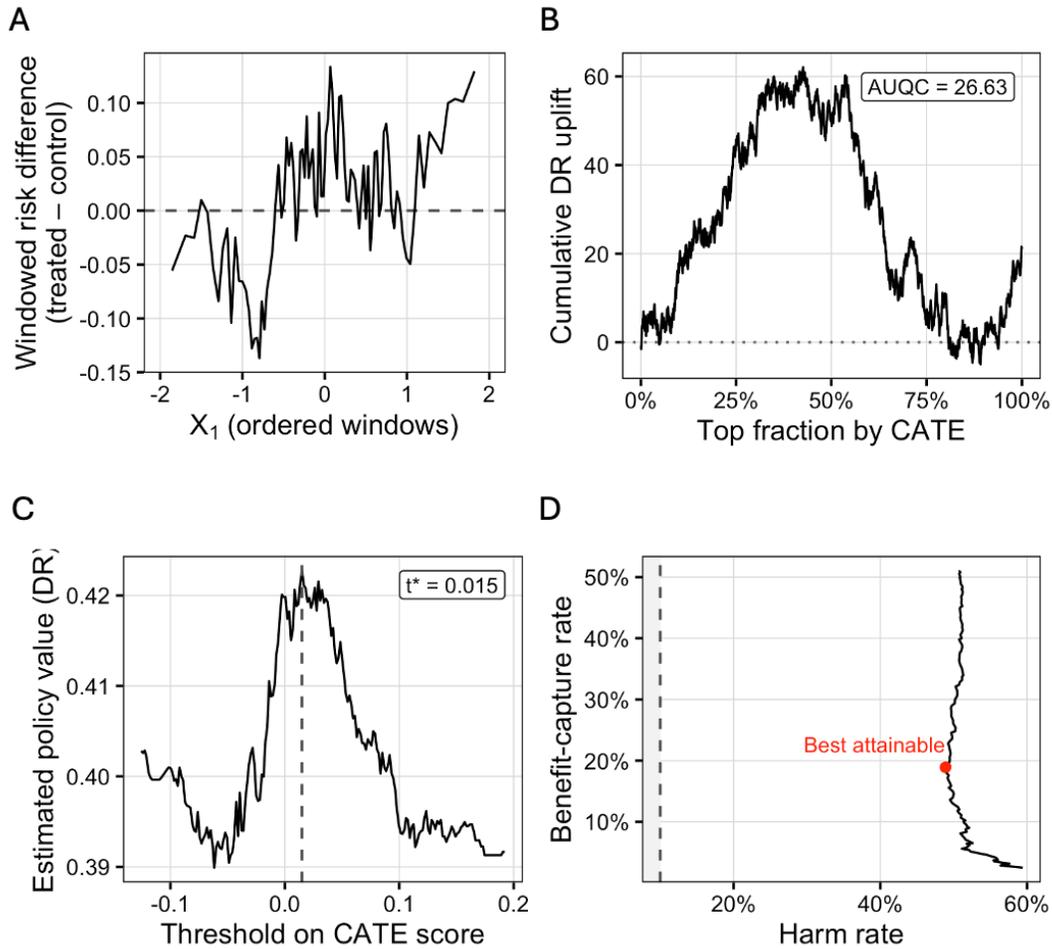

**Supplementary Figure S2. Results from one Weak HTE replicate applying the two-stage workflow.** (A) STEPP exploration of treatment effect heterogeneity along biomarker $X_1$. The curve starts below zero with a gentle increase at larger $X_1$, indicating weak but nonnegligible effect modification. (B) The curve rises and then falls (AUQC = 26.63), reflecting that treatment benefit is concentrated among top ranked patients but with less separation than in the Strong HTE scenario. (C) Cross fitted DR policy value $\hat{V}(\pi_t)$ as a function of the CATE threshold $t$. The curve shows an interior maximum at $t^* = 0.015$, showing that an individualized rule that treats only patients with larger estimated CATE yields a small but positive improvement in outcome probability relative to treating all or none. (D) NP rule summarizing harm-benefit tradeoffs across thresholds. No threshold satisfies the 10% harm constraint.